\documentclass[journal]{IEEEtran}

\usepackage{float}  

\usepackage{caption}
\usepackage{subcaption}

\usepackage{stfloats}

\usepackage{algorithm}  
\usepackage{algpseudocode}  
\usepackage{amsmath}  

\usepackage{graphicx}  
\usepackage{color}
\usepackage{cite}

\usepackage{graphicx}  
\usepackage{epstopdf}

\usepackage{graphicx}  

\usepackage{multirow,booktabs} 
\usepackage{makecell} 
\usepackage{diagbox}

\usepackage{stfloats}
\usepackage{graphicx}

\usepackage{url}
\usepackage[hidelinks]{hyperref}

\ifCLASSINFOpdf
\else
\fi
\usepackage{amsthm,amsmath,amssymb}

\begin{document}
	
\captionsetup[figure]{labelfont={},labelformat={default},labelsep=period,name={Fig.}}


\captionsetup[table]{labelfont={},labelformat={default},labelsep=period,name={Table}}

%
\title{Selective Depthwise Separable Convolution for Lightweight Joint Source-Channel Coding in Wireless Image Transmission}








\author{Ming~Ye,~\IEEEmembership{Member,~IEEE,}~Kui Cai,~\IEEEmembership{Senior Member,~IEEE,}~Cunhua Pan,~\IEEEmembership{Senior Member,~IEEE,} Zhen Mei,~\IEEEmembership{Member,~IEEE,} Wanting Yang, and Chunguo Li,~\IEEEmembership{Senior Member,~IEEE}

\thanks{

Manuscript received 22 April 2026; revised 17 July 2026; accepted
22 September 2026. 
This work was supported in part by the Joint SMU-SUTD Grant 24-SIS-SMU-052, in part by A*STAR under its RIE2025 Innovation \& Enterprise Industry Alignment Fund -- Industry Collaboration Projects (Award I2501E0064, Grant H25-MCP3438), in part by the National Natural Science Foundation of China under Grants U25A20393, U25B2007, and 62331023, and in part by the National Key R\&D Program of China under Grant 2024YFC3807900.
(\textit{Corresponding author: Kui Cai}.)

 	
M. Ye, K. Cai, and W. Yang are with the Singapore University of Technology and Design, Singapore 487372 (e-mail: 230198460@aa.seu.edu.cn; cai$_{-}$kui@sutd.edu.sg; wanting$_{-}$yang@sutd.edu.sg).

C. Pan and C. Li are with the National Mobile Communications Research Laboratory, Southeast University, Nanjing 210096, China (e-mail: cpan@seu.edu.cn;  chunguoli@seu.edu.cn).

Z. Mei is with
the School of Electronic and Optical Engineering, Nanjing University of Science and Technology, Nanjing 210094, China (e-mail: meizhen@njust.edu.cn).
The source code is publicly available on
\href{https://github.com/Yeming198460/Selective-Depthwise-Separable-Convolution-for-Lightweight-Joint-Source-Channel-Coding2-}{\textbf{GitHub}}.


}

}



%
%

\markboth{}%
{Shell \MakeLowercase{\textit{et al.}}: Bare Demo of IEEEtran.cls for IEEE Journals}
%



\maketitle

\begin{abstract}

Depthwise separable convolutional (DSConv) layers have been widely adopted to replace convolutional (Conv) layers in deep learning (DL)-based joint source-channel coding (JSCC) schemes to reduce computational complexity.
However, a systematic investigation of Conv-to-DSConv replacement at different layer positions and ratios in DL-based JSCC systems for wireless image transmission remains largely unexplored.
In this letter, we propose a  lightweight convolutional neural network (CNN)-based JSCC framework with a configurable and complexity-aware Conv-to-DSConv replacement strategy, enabling flexible replacement at different layer positions and ratios.
By varying the replacement ratio, we obtain models with different computational complexities and analyze their impact on reconstruction performance.
Furthermore, we investigate how Conv-to-DSConv replacement at different layer positions influences reconstruction quality under a fixed replacement ratio.
Our results show that Conv-to-DSConv replacement at the intermediate layers of the encoder and decoder achieves a favorable complexity-performance trade-off, providing practical design guidelines for lightweight CNN-based JSCC systems.
Extensive experiments demonstrate that the proposed framework achieves substantial reductions in computational complexity with only a slight performance degradation, enabling flexible complexity-performance trade-offs for resource-constrained edge devices.

\end{abstract}



%
\IEEEpeerreviewmaketitle

\section{Introduction}
%
%
%
%

\IEEEPARstart{S}{emantic} communication has emerged as a promising technology for sixth-generation (6G) communication systems, serving as an innovative paradigm for integrating communications and artificial intelligence \cite{r1}.
Unlike traditional communication systems, it focuses on conveying the meaning of transmitted information instead of solely minimizing bit error rates. 
Deep learning (DL)-based joint source-channel coding (JSCC) has become a promising and widely adopted enabler of this paradigm \cite{r2, rb2}. It directly maps the original information into continuous channel inputs, facilitating end-to-end optimization of the communication system \cite{r3, r4}.


Recent advancements in DL, especially in image processing \cite{r5, r6} and natural language processing \cite{r7}, 
have further accelerated the development of semantic communication.
Nevertheless, the limited computational, memory, and energy resources of edge devices pose significant challenges to the deployment of DL-based JSCC models.
Reducing the model size of DL-based JSCC systems is therefore essential, as it directly affects real-time performance, storage requirements, energy consumption, and deployment flexibility. Consequently, there is a pressing need for lightweight DL-based JSCC methods that achieve acceptable performance with low computational complexity.

Several recent studies have explored lightweight DL-based JSCC approaches \cite{r9, r10, r11, r12, r13}. 
To reduce computational overhead, the authors of \cite{ra8} proposed a JSCC method based on a state-space model architecture for wireless image transmission.
A JSCC framework based on a Swin Transformer backbone was proposed in \cite{ra9}, achieving lower computational complexity than conventional Transformer-based models.
However, both methods still incur high computational complexity, which limits their deployment on resource-constrained edge devices.
In \cite{r9}, a lightweight method for semantic image reconstruction and classification was proposed, which employs ConvNeXt-based modules with depthwise separable convolutional (DSConv) layers to reduce computational complexity.
Similarly, the authors of \cite{r10} proposed a lightweight DL-based JSCC approach, termed DeepJSCC-T, which uses ConvNeXt-based modules to reduce model complexity for wireless image transmission.
A modified MobileNetV2 comprising DSConv layers was used in \cite{r11} to reduce computational overhead in the task-oriented JSCC scheme for downstream tasks such as face detection and image classification.  
A lightweight DL-based JSCC framework with an adaptive module was proposed in \cite{r12}, where convolutional (Conv) layers were replaced with DSConv layers to reduce model complexity for feature extraction.     
The authors of \cite{r13} proposed a lightweight JSCC method for image reconstruction, where DSConv layers and a mixture block were used to reduce complexity.     
However, the aforementioned DL-based JSCC works \cite{r9,r10,r11,r12,r13} mainly focus on developing specific lightweight architectures, while a systematic investigation of Conv-to-DSConv replacement at different layer positions and ratios in JSCC systems for wireless image transmission is still lacking.




To address these limitations, we propose a lightweight convolutional neural network (CNN)-based JSCC framework, termed DSC-JSCC, for wireless image transmission.
The framework incorporates a configurable and complexity-aware Conv-to-DSConv replacement strategy that enables flexible replacement at different layer positions and ratios.
Guided by the layer-wise distribution of floating-point operations (FLOPs), we selectively replace the Conv layers in the encoder and the transposed convolutional (TConv) layers in the decoder with their DSConv and depthwise separable transposed convolutional (DSTConv) counterparts, respectively, to reduce computational complexity.
The primary contributions of this letter are summarized as follows:

$\bullet$ 
We propose DSC-JSCC, a lightweight CNN-based JSCC framework with a configurable and complexity-aware Conv-to-DSConv replacement strategy guided by the layer-wise FLOP distribution.







$\bullet$ We systematically investigate the impact of Conv-to-DSConv replacement at different layer positions and ratios on performance and complexity. 
Our results provide practical design guidelines for lightweight CNN-based JSCC systems and
show that Conv-to-DSConv replacement at the intermediate layers of the encoder and decoder achieves a favorable complexity-performance trade-off.



$\bullet$ Extensive experiments demonstrate that the proposed framework achieves significant parameter reduction with only slight performance degradation, enabling flexible complexity-performance trade-offs for resource-constrained edge devices.

\section{System  Model}

\begin{figure}
	\centering
	\includegraphics[width=7.5cm]{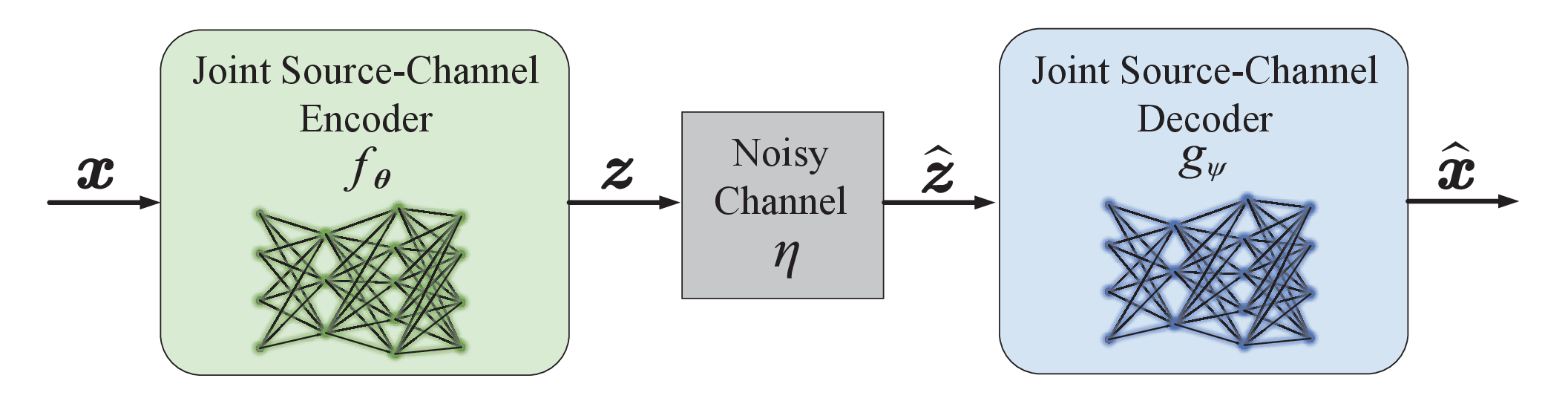}
	\caption{Block diagram of the image transmission system for the proposed framework.}
	\label{tex10_fig1}
\end{figure}

We consider an end-to-end DL-based JSCC transmission system consisting of a trainable encoder $f_{\boldsymbol{\theta}}$, a trainable decoder $g_{\boldsymbol{\psi}}$, and a noisy wireless channel, as shown in Fig.~\ref{tex10_fig1}. 
$\boldsymbol{\theta}$ and $\boldsymbol{\psi}$ denote the parameters of the encoder and the decoder, respectively.
The source image $\boldsymbol{x}\in\mathbb{R}^d$ is mapped into a complex-valued vector $\boldsymbol{z}\in\mathbb{C}^k$ by the encoder as
\begin{equation}
	\boldsymbol{z}=f_{\boldsymbol{\theta}}\left(\boldsymbol{x},\rho\right)\in\mathbb{C}^k,
	\label{encoding}
\end{equation}
where $\rho$ is the bandwidth compression ratio defined as $\rho=k/d$.
After encoding, the vector $\boldsymbol{z}$ is transmitted over a Rayleigh slow-fading channel, yielding
\begin{equation}
	\hat{\boldsymbol{z}}
	=
	\eta(\boldsymbol{z},\sigma^2)
	=
	h\boldsymbol{z}
	+
	\boldsymbol{n},
	\label{transfer_function}
\end{equation}
where $h\sim\mathcal{CN}(0,1)$ denotes the Rayleigh fading channel coefficient, $\eta(\cdot)$ denotes the channel transfer function, and $\boldsymbol{n}\sim\mathcal{CN}(\mathbf{0},\sigma^2\mathbf{I})$ denotes circularly symmetric complex Gaussian noise, with $\sigma^2$ representing the noise power.
The transfer function $\eta(\cdot)$ can be extended to other channel models,  as long as it remains differentiable \cite{r3, r4}.
Note that the vector $\boldsymbol{z}$ is obtained by reshaping the output of the final convolutional layer of the encoder into a complex-valued vector $\tilde{\boldsymbol{z}}$, which is then normalized to satisfy the average transmit power constraint $\tilde{P}$ as
$\boldsymbol{z}=\sqrt{k\tilde{P}}\frac{\tilde{\boldsymbol{z}}}{\sqrt{\tilde{\boldsymbol{z}}^H\tilde{\boldsymbol{z}}}}$.
The average transmit power is normalized to $\tilde{P}=1$, and the signal-to-noise ratio (SNR) is defined as
	$\mathrm{SNR}=10\log_{10}(\tilde{P}/\sigma^{2})$.

Finally, the corrupted vector $\hat{\boldsymbol{z}}$ is fed into the decoder to reconstruct the image as
\begin{equation}
	\hat{\boldsymbol{x}}=g_{\boldsymbol{\psi}}\left(\hat{\boldsymbol{z}}\right)\in\mathbb{R}^d.
	\label{decoding}
\end{equation}

Unlike existing lightweight DL-based JSCC methods that mainly focus on designing specific lightweight architectures, the proposed framework adopts a configurable and complexity-aware Conv-to-DSConv replacement strategy. This strategy is guided by the layer-wise FLOP distribution, enabling flexible complexity-performance trade-offs for resource-constrained deployment.



\section{Proposed Framework}




\subsection{Lightweight DSC-JSCC Architecture}

\begin{figure}
	\centering
	\includegraphics[width=7.0cm]{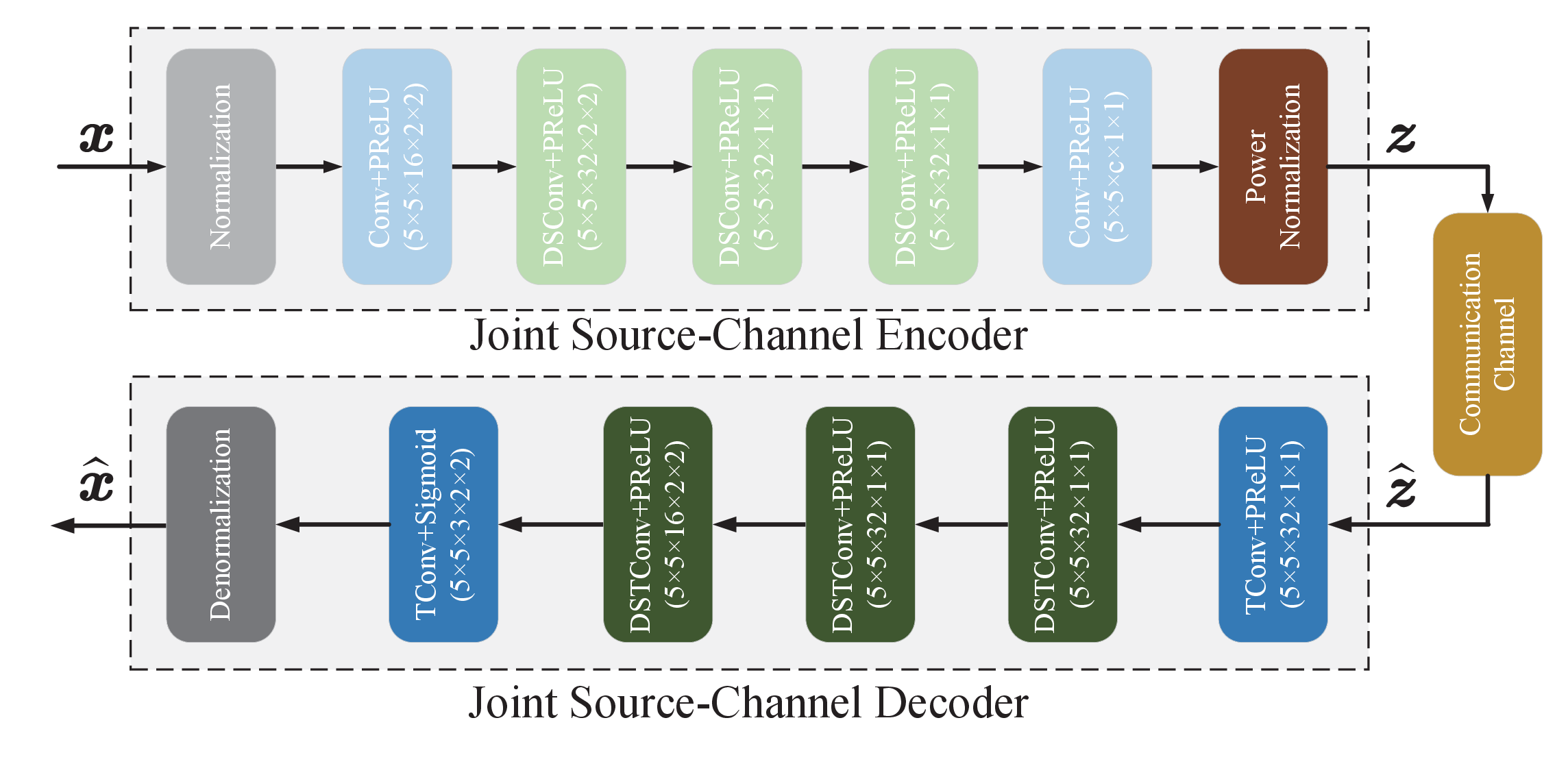}
	\caption{Architecture of the proposed DSC-JSCC-60 (E2D2) model.}
	\label{tex10_Architecture_DSCJSCC_60}
	\vspace{-4mm}
\end{figure}


In this work, we propose a lightweight CNN-based JSCC framework, termed DSC-JSCC, with a configurable and complexity-aware Conv-to-DSConv replacement strategy. The strategy enables flexible replacement at different layer positions and ratios. Specifically, we selectively replace the Conv layers in the encoder and the TConv layers in the decoder with DSConv layers and DSTConv layers, respectively. The architecture of DSC-JSCC-60 (E2D2), a representative variant of the proposed framework, is illustrated in Fig. \ref{tex10_Architecture_DSCJSCC_60}. Each DSConv layer consists of a depthwise convolution for spatial feature extraction followed by a pointwise convolution for channel mixing. Each DSTConv layer consists of a transposed depthwise convolution for spatial upsampling followed by a pointwise convolution for channel mixing.

The normalization layer scales the pixel values of the encoder input $\boldsymbol{x}$ from [0, 255] to [0, 1] to mitigate potential exploding gradients. As shown in Fig. \ref{tex10_Architecture_DSCJSCC_60}, the Conv and DSConv layers in the joint source-channel encoder extract features from the normalized input.
Compared with Conv layers, DSConv layers significantly reduce computational complexity while maintaining efficient feature extraction. The extracted features are transformed into the complex-valued vector $\boldsymbol{z}$ using a power normalization layer to satisfy the average transmit power constraint. 
Note that we can adjust the number of channels $c$ to vary the bandwidth compression ratio.

The resulting vector $\boldsymbol{z}$ is transmitted over the wireless channel as defined in \eqref{transfer_function}, yielding the corrupted vector $\hat{\boldsymbol{z}}$. The corrupted vector is then reshaped into a real-valued matrix and processed by the joint source-channel decoder.
In the decoder, TConv and DSTConv layers extract features from the reshaped matrix. Similar to DSConv layers, DSTConv layers significantly reduce computational complexity while maintaining efficient feature extraction. Finally, the denormalization layer scales the decoder output to the [0, 255] range, producing the final reconstructed image $\hat{\boldsymbol{x}}$.


The proposed framework consists of two phases: the offline training phase and the online reconstruction phase. During the offline training phase, the encoder and the decoder of the proposed framework are optimized end-to-end by minimizing the mean squared error (MSE) loss. During the online reconstruction phase, a test image $\boldsymbol{x}$ is fed into the trained framework to obtain the reconstructed image:
\begin{equation}
	\hat{\boldsymbol{x}} = g_{\boldsymbol{\psi}}(\hat{\boldsymbol{z}}) = g_{\boldsymbol{\psi}}\big(\eta(f_{\boldsymbol{\theta}}(\boldsymbol{x}, \rho), \sigma^2)\big).
\end{equation}



\begin{table}[htbp]
	
	\centering
	\caption{Architectures of DSC-JSCC variants with different replacement ratios.}
	
	\scalebox{1}
	{
		\begin{tabular}{c|c|c}
			
			\toprule[0.8pt]
			Model	& Encoder  &	Decoder     \\   \hline  
			
			DSC-JSCC-20	& \makecell{DSConv (Layer 1) \\ Conv (Layers 2--5)} &	\makecell{DSTConv (Layer 1) \\ TConv (Layers 2--5)} \\ \hline
			
			DSC-JSCC-40	& \makecell{DSConv (Layers 1--2) \\ Conv (Layers 3--5)}     &\makecell{DSTConv (Layers 1--2) \\ TConv (Layers 3--5)} \\ \hline
			
			\makecell{DSC-JSCC-60 \\ (E1D1)} 	& \makecell{DSConv (Layers 1--3) \\ Conv (Layers 4--5)}  &	\makecell{DSTConv (Layers 1--3) \\ TConv (Layers 4--5)} \\  \hline  
			DSC-JSCC-80	& \makecell{DSConv (Layers 1--4) \\ Conv (Layer 5)}   &	\makecell{DSTConv (Layers 1--4) \\ TConv (Layer 5)}   \\ \hline

			DSC-JSCC-100	&  DSConv (Layers 1--5)  &	DSTConv (Layers 1--5)    \\

			\bottomrule[0.8pt]
		\end{tabular}	
	}
	\label{DSC_ratios}
\end{table}  

\vspace{-3mm}

\subsection{Layer Replacement at Different Ratios} 

The proposed DSC-JSCC framework adopts a 
configurable and complexity-aware Conv-to-DSConv replacement strategy.
Based on the strategy, various DSC-JSCC variants are obtained by replacing different ratios of Conv layers in the encoder and TConv layers in the decoder with DSConv and DSTConv layers, respectively, while keeping all other settings unchanged.  
To evaluate the impact of different replacement ratios, the early Conv layers in the encoder and the early TConv layers in the decoder are replaced with their corresponding DSConv and DSTConv layers. The structures of these ratio-based variants are summarized in Table \ref{DSC_ratios}. 
As shown in Table \ref{DSC_ratios}, DSC-JSCC-$X$ denotes the variant in which $X\%$ of Conv and TConv layers are replaced by DSConv and DSTConv layers, respectively, where $X \in \{20, 40, 60, 80, 100\}$. 
A higher value of $X$ corresponds to a DSC-JSCC variant with lower computational complexity.




\vspace{-1mm}

\subsection{Layer Replacement at Different Positions}

\begin{table}[htbp]
	
	\centering
	\scriptsize
	\caption{Architectures of DSC-JSCC variants with different replacement positions at a $60\%$ replacement ratio.}

	\scalebox{1}
	{
		\begin{tabular}{c|c|c}
			
			\toprule[0.8pt]
			Model	& Encoder  &	Decoder     \\   \hline  
			
DSC-JSCC-60 (E2D1)	& \makecell{Conv (Layer 1) \\ DSConv (Layers 2--4)\\ Conv (Layer 5)} &	\makecell{DSTConv (Layers 1--3) \\ TConv (Layers 4--5)} \\  \hline

DSC-JSCC-60 (E2D2)	& \makecell{Conv (Layer 1) \\ DSConv (Layers 2--4)\\ Conv (Layer 5)} &	\makecell{TConv (Layer 1) \\ DSTConv (Layers 2--4)\\ TConv (Layer 5)}  \\ \hline

DSC-JSCC-60 (E2D3)	& \makecell{Conv (Layer 1) \\ DSConv (Layers 2--4)\\ Conv (Layer 5)}  &	\makecell{TConv (Layers 1--2) \\ DSTConv (Layers 3--5)}   \\ \hline

DSC-JSCC-60 (E1D2)	&  \makecell{DSConv (Layers 1--3) \\ Conv (Layers 4--5)}   &	\makecell{TConv (Layer 1) \\ DSTConv (Layers 2--4)\\ TConv (Layer 5)}    \\    \hline

DSC-JSCC-60 (E3D2)	&  \makecell{Conv (Layers 1--2) \\ DSConv (Layers 3--5)}   &\makecell{TConv (Layer 1) \\ DSTConv (Layers 2--4)\\ TConv (Layer 5)}  \\  
			
\bottomrule[0.8pt]
		\end{tabular}	
	}
	\label{DSC_positions} 
	 \vspace{-2.5mm} 
\end{table}  

 

Let E1, E2, and E3 denote the early, middle, and late Conv layers in the encoder, respectively, while D1, D2, and D3 denote the early, middle, and late TConv layers in the decoder, respectively.
Accordingly, DSC-JSCC-60 (E2D2) denotes the variant in which E2 and D2 are replaced with DSConv and DSTConv layers, respectively.
Similarly, DSC-JSCC-60 (E1D2) denotes the variant in which E1 and D2 are replaced with DSConv and DSTConv layers, respectively.
The FLOPs of each layer are primarily determined by the feature-map resolution and the number of input and output channels, since all layers share the same kernel size. In our encoder-decoder architecture, the intermediate layers (i.e., E2 and D2) account for the largest proportion of the overall FLOPs. Thus, prioritizing the replacement of these intermediate layers with DSConv and DSTConv layers, respectively, yields greater reductions in FLOPs than replacing the early or late layers. This replacement strategy, guided by the layer-wise FLOP distribution, is complexity-aware rather than based on empirical trial and error.


Based on the replacement strategy, additional DSC-JSCC variants at a fixed replacement ratio are generated by replacing Conv and TConv layers at different positions in the encoder and the decoder with DSConv and DSTConv layers, respectively. Meanwhile, other settings remain unchanged. To evaluate the effect of layer positions, the early, middle, and late Conv layers in the encoder and the TConv layers in the decoder are selectively replaced at a $60\%$ replacement ratio.
Table \ref{DSC_positions} summarizes the structures of position-based DSC-JSCC variants. 
DSC-JSCC variants include models with different replacement ratios (e.g., DSC-JSCC-20 and DSC-JSCC-40) and different layer positions (e.g., DSC-JSCC-60 (E1D1) and DSC-JSCC-60 (E2D2)).  
Among these variants, DSC-JSCC-60 (E2D2) is selected as the default configuration due to its favorable complexity-performance trade-off.







\begin{figure*}[htbp]  
\vspace{-3mm}
	\begin{minipage}[t]{0.49\textwidth}
		
		\centering	
		\includegraphics[width=5.8cm]{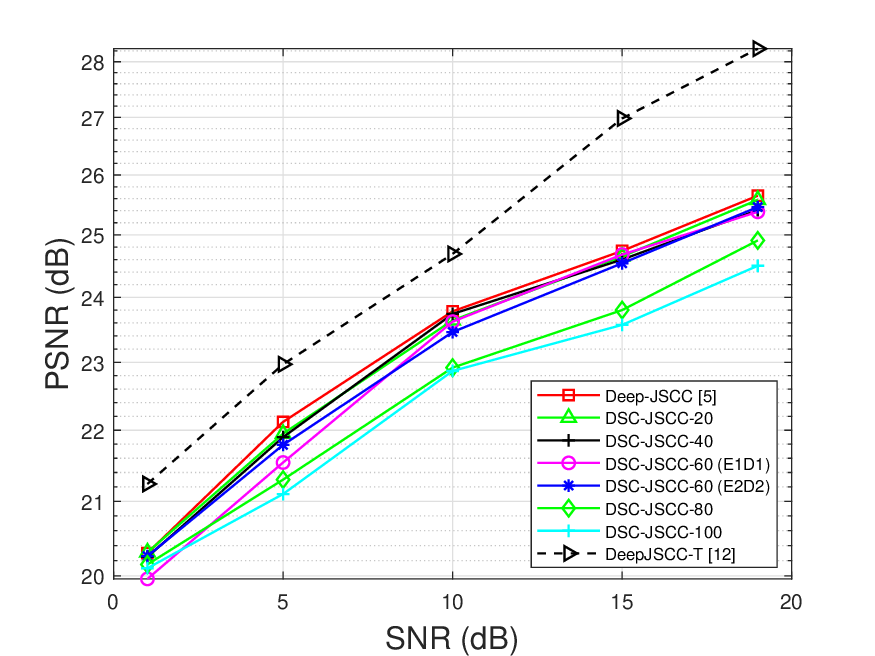} 
		\subcaption{}  
	\end{minipage}	
	\begin{minipage}[t]{0.49\textwidth}			
		\centering	
		\includegraphics[width=5.8cm]{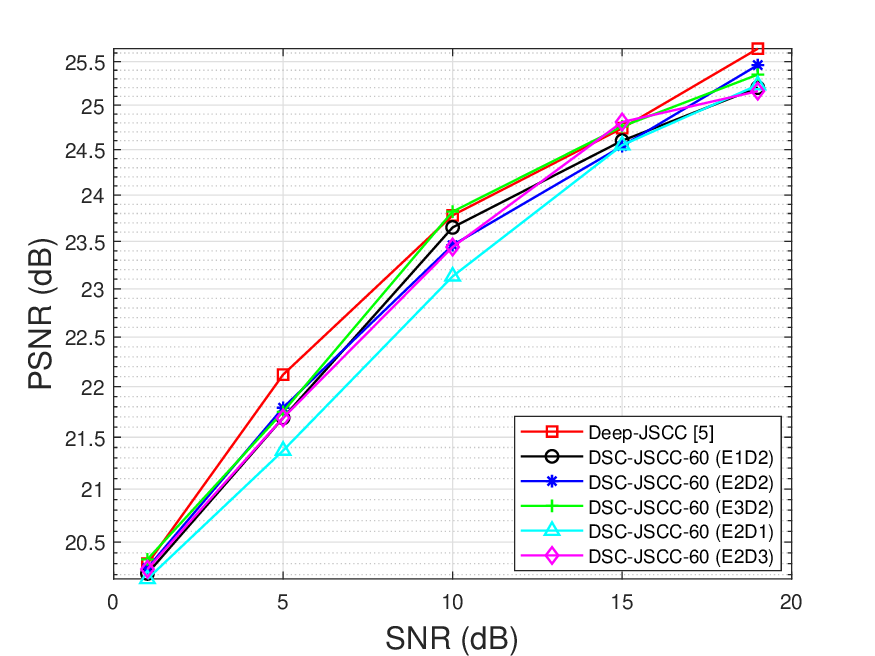} 
		\subcaption{}
	\end{minipage}	
	
	\caption{PSNR comparison of DSC-JSCC variants and baseline JSCC models under different replacement ratios and positions on the CIFAR-10 dataset.}	  
	\label{tex10_cHQ_ratios_psnr/lpips}	
	\vspace{-3mm}	
\end{figure*}

\begin{figure*}[htbp]  
\vspace{-2mm}
	\begin{minipage}[t]{0.49\textwidth}
		
		\centering	
		\includegraphics[width=5.8cm]{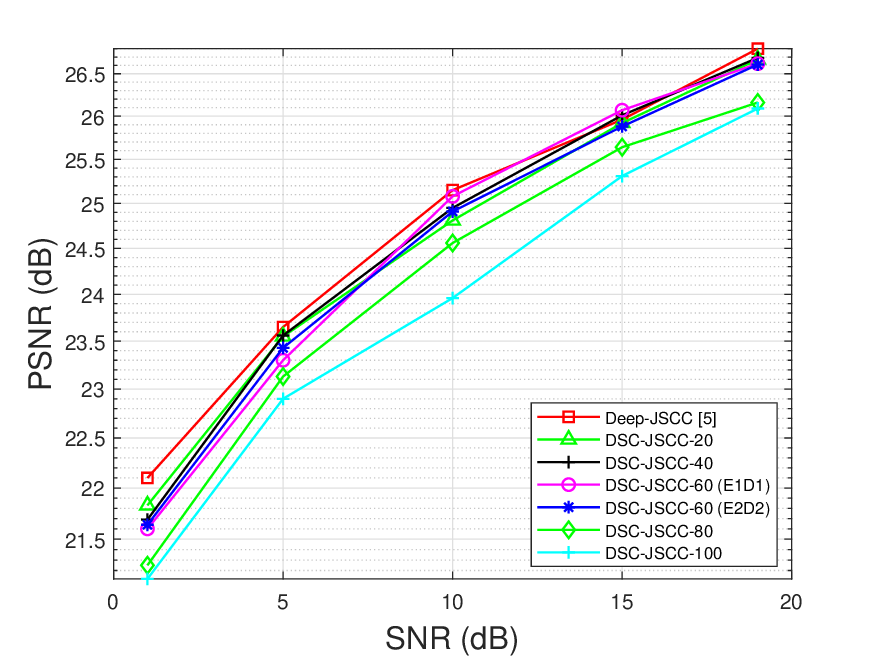} 
		\subcaption{}
	\end{minipage}	
	\begin{minipage}[t]{0.49\textwidth}			
		\centering	
		\includegraphics[width=5.8cm]{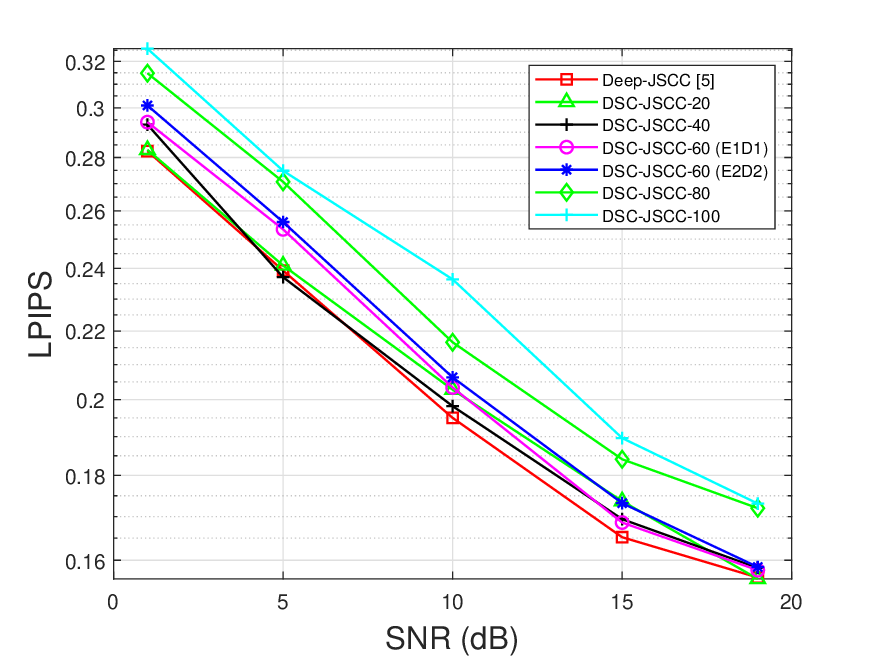} 
		\subcaption{}  
	\end{minipage}	
	
	\caption{PSNR and LPIPS performance of DSC-JSCC variants with different replacement ratios on the CelebA dataset.}
	
	\label{tex10_celebA_ratios_psnr/lpips}		
	\vspace{-4mm}
\end{figure*}

\section{Numerical Results}



To evaluate the performance of the DSC-JSCC variants, we compare them with Deep-JSCC \cite{r3}, a relevant CNN-based baseline, and DeepJSCC-T \cite{r10}, a representative state-of-the-art reference, using the CIFAR-10 \cite{rb13} and CelebA \cite{r14} datasets.
The CIFAR-10 dataset contains 50,000 training images and 10,000 test images, where each image has a resolution of $32\times32\times3$. The CelebA dataset contains 160,000 training images and 10,000 test images, where each image is center-cropped and resized to $64\times64\times3$.
The quality of the reconstructed images is evaluated using the peak signal-to-noise ratio (PSNR) and the learned perceptual image patch similarity (LPIPS) \cite{r15}. LPIPS is computed using a pre-trained VGG network.
The Adam optimizer is used for all DSC-JSCC variants, with an initial learning rate of 0.001. The batch size is set to 32, and the models are trained for 100 epochs.
Unless otherwise specified, all experiments are conducted under a Rayleigh slow-fading channel with $\rho=1/12$.
For a fair performance comparison, all compared models are trained separately at each SNR value and evaluated under the corresponding SNR condition.
All models are implemented in Python 3.8.20 using TensorFlow 2.6.0.




To achieve the desired bandwidth compression ratio $\rho$, the key parameters $k$ and $c$ are determined as follows.  
The number of source symbols is given by $d =  W \times H \times C$, where $W$, $H$, and $C$ denote the width, height, and number of channels of the input image, respectively.  
The number of transmitted symbols is given by $k=\lfloor \rho \times d \rfloor$, where $\lfloor \cdot \rfloor$ denotes the floor operator.  
The number of output channels $c$ is computed as  
$c=\lfloor 2k / \left( \bar{H}\times \bar{W} \right) \rfloor$,  
where $\bar{H}\times \bar{W}$ represents the spatial dimensions of the feature map produced by the final Conv layer in the encoder.



Fig.~\ref{tex10_cHQ_ratios_psnr/lpips}(a) presents the PSNR results on the CIFAR-10 dataset. The PSNR of the DSC-JSCC variants generally decreases as the replacement ratio increases. Nevertheless, both DSC-JSCC-20 and DSC-JSCC-40 achieve PSNR values comparable to those of Deep-JSCC while exhibiting lower computational complexity. DSC-JSCC-60 (E2D2) exhibits slightly lower PSNR than DSC-JSCC-20 while further reducing computational complexity, with only a slight PSNR degradation compared with Deep-JSCC.
Furthermore, DSC-JSCC-60 (E2D2) achieves performance comparable to that of DSC-JSCC-60 (E1D1) with lower complexity. Although DSC-JSCC-80 achieves lower complexity than DSC-JSCC-60 (E2D2), it suffers from noticeable performance degradation. DeepJSCC-T achieves higher PSNR than DSC-JSCC-60 (E2D2). However, it incurs substantially higher computational complexity, making it less suitable for resource-constrained edge deployment. DeepJSCC-T is included only as a representative state-of-the-art reference with substantially higher computational complexity, rather than as a complexity-equivalent baseline.



\vspace{-0.45mm}



As shown in Fig.~\ref{tex10_cHQ_ratios_psnr/lpips}(b), different replacement positions lead to different PSNR values for the DSC-JSCC-60 variants. The performance gap between the DSC-JSCC variants and Deep-JSCC remains small, indicating that substantial complexity reductions can be achieved with only a slight performance degradation.
At a 60\% replacement ratio, DSC-JSCC-60 (E2D2) achieves performance comparable to that of the other variants, while exhibiting the lowest computational complexity among them. 
Since the intermediate layers of the encoder and decoder account for the largest proportion of the overall FLOPs, replacing these intermediate layers yields greater reductions in computational complexity than replacing the early or late layers. Thus, DSC-JSCC-60 (E2D2) achieves a favorable complexity-performance trade-off, making it well-suited for resource-constrained edge deployment.



We further evaluate the impact of different replacement ratios on the PSNR and LPIPS performance of the DSC-JSCC variants on the CelebA dataset.
As shown in Fig.~\ref{tex10_celebA_ratios_psnr/lpips}(a), DSC-JSCC-20, DSC-JSCC-40, and both DSC-JSCC-60 variants achieve comparable PSNR. Among these models, DSC-JSCC-60 (E2D2) exhibits the lowest computational complexity and thus provides a favorable complexity-performance trade-off. Although DSC-JSCC-80 and DSC-JSCC-100 further reduce complexity, they suffer from noticeable performance degradation compared with the other DSC-JSCC variants.  
A similar trend is observed in Fig.~\ref{tex10_celebA_ratios_psnr/lpips}(b), indicating that DSC-JSCC-60 (E2D2) also achieves a favorable trade-off between LPIPS and computational complexity.
Overall, DSC-JSCC-60 (E2D2) achieves a favorable complexity-performance trade-off on the CelebA dataset.


\begin{table}[htbp]
	   \vspace{-1.5mm}
	\centering
	\caption{Computational complexity comparison of DSC-JSCC variants against existing DL-based JSCC models.}

	\scalebox{1}
	{
		\begin{tabular}{c|c|c}
			
			\toprule[0.8pt]
			Model	& Parameters (K)  &	FLOPs (M) 
			\\   \hline  
			

DeepJSCC-T	& 4670.2  & 206.8 \\ \hline 

Deep-JSCC	& 164.4 & 19.3 \\ \hline  

DSC-JSCC-20	&  157.4 &	18.0    \\ \hline  

DSC-JSCC-40	&  121.8 &	13.5  \\ \hline 

DSC-JSCC-60 (E1D1) 	& 74.3 &7.4  \\ \hline  

DSC-JSCC-60 (E2D1) 	& 51.6 &  4.9 \\ \hline	 

\makecell{DSC-JSCC-60 (E2D2)  \\ } 	&  \textbf{46.1}  & \textbf{4.7}  \\ \hline

DSC-JSCC-60 (E2D3) 	& 69.1  &  8.1 \\ \hline 	

DSC-JSCC-60 (E1D2) 	& 68.8  &  7.2 \\ \hline 	 

\makecell{DSC-JSCC-60 (E3D2)  \\ } 	& 52.7   & 5.5 \\ \hline	

DSC-JSCC-80	&  39.1 &	3.4   \\ \hline  

DSC-JSCC-100 &  \textbf{33.1} &	\textbf{3.0}  \\    

			\bottomrule[0.8pt]
		\end{tabular}	
	}
	\label{DSC_complexity}
\vspace{-1mm}
\end{table}

We evaluate the number of parameters and FLOPs of the DSC-JSCC variants and baseline models on the CIFAR-10 dataset with a batch size of 1. 
As shown in Table \ref{DSC_complexity}, DeepJSCC-T exhibits the highest computational complexity among all models, with substantially higher complexity than Deep-JSCC.
Specifically, compared with Deep-JSCC, DeepJSCC-T requires approximately 28.4 times as many parameters and 10.7 times as many FLOPs.
In contrast, DSC-JSCC-60 (E2D2) reduces the number of parameters and FLOPs by 72.0\% and 75.6\%, respectively, compared with Deep-JSCC, while incurring only a slight performance degradation.
These reductions result from replacing 60\% of the Conv and TConv layers with their DSConv and DSTConv counterparts, respectively. This replacement reduces the computational cost by decomposing standard convolutions into depthwise and pointwise convolutions \cite{r16}.
Among the DSC-JSCC-60 variants with a fixed replacement ratio of 60\%, DSC-JSCC-60 (E2D2) achieves the lowest computational complexity. It further reduces the number of parameters and FLOPs by 38.0\% and 36.5\%, respectively, compared with DSC-JSCC-60 (E1D1). Although DSC-JSCC-80 and DSC-JSCC-100 achieve lower computational complexity than DSC-JSCC-60 (E2D2), they suffer from significant performance degradation. Overall, DSC-JSCC-60 (E2D2) achieves a favorable complexity-performance trade-off, making it attractive for practical deployment in resource-constrained scenarios.

\section{Conclusion}



In this letter, we proposed a lightweight CNN-based JSCC framework. It employs a configurable and complexity-aware Conv-to-DSConv replacement strategy guided by the layer-wise FLOP distribution. The replacement strategy enables replacement at different ratios and layer positions, affecting both performance and computational complexity.
Our results demonstrate that Conv-to-DSConv replacement at the intermediate layers of the encoder and decoder achieves a favorable complexity-performance trade-off, providing practical design guidelines for lightweight CNN-based JSCC systems. Our framework achieves substantial reductions in parameters and FLOPs with only a slight performance degradation, making it well suited for resource-constrained edge deployment.

\vspace{12pt}

\ifCLASSOPTIONcaptionsoff
  \newpage
\fi



%


\end{document}